# Specific Heat of Zn-Doped YBa$_2$Cu$_3$O$_{6.95}$: Possible Evidence for Kondo Screening in the Superconducting State

D. L. Sisson, S. G. Doettinger, and A. Kapitulnik

*Departments of Applied Physics and of Physics, Stanford University, Stanford, CA 94305, USA*

R. Liang, D. A. Bonn, and W. N. Hardy

*Department of Physics, UBC, Vancouver, British Columbia V6T1Z1*

(August 12, 2018)

The magnetic field dependence of the specific heat of Zn-doped single crystals of YBa$_2$Cu$_3$O$_{6.95}$ was measured between 2 and 10 K and up to 8 Tesla. Doping levels of 0, 0.15%, 0.31%, and 1% were studied and compared. In particular we searched for the Schottky anomaly associated with the Zn-induced magnetic moments.

PACS numbers: 74.25, 74.62.D, 75.20.H

Among systems for controlled studies of the effect of disorder on the properties of cuprate superconductors, Zn-doped YBa$_2$Cu$_3$O$_{7-\delta}$ (YBCO) is a particularly interesting choice. Unlike most dopants, Zn substitutes for Cu on the planes [1] and causes significant changes to the local electronic structure without much change of hole concentration [2]. In the superconducting state Zn substitution dramatically changes both the transition temperature [3] and the temperature dependence of the low-temperature penetration depth [4]. NMR and magnetization measurements in the normal state suggested that Zn induces a weak magnetic moment $\mu_{eff}$=0.32 to 0.36 $\mu_B$/Zn for fully oxygenated crystals, and that this moment increases as the oxygen content decreases [3,5]. An oxygen reduced sample (O$_{6.66}$) which exhibited a spin gap showed a moment of $0.86\mu_B$/Zn [6]. A model in which each Zn atom effects the local magnetic order of nearby copper atoms on the scale of the magnetic copper correlation length (which varies with $\delta$) explains the normal-state magnetic measurements on this system [7]. In the superconducting state, NMR measurements suggested that in Zn substituted YBCO there exists a finite electronic density of states (DOS) at the Fermi level [8]. Early specific heat measurements [9] seemed to agreed with this conclusion indicating increased DOS in the superconducting state in Zn substituted samples.

Simple valence counting suggests that if the Zn impurity maintains a nominal Cu$^{2+}$ charge, the Zn$^{2+}$ would have a $(3d)^{10}$, $S$=0 configuration and act as a nonmagnetic impurity. Thus, it is important to understand whether a nonmagnetic impurity can induce a local moment on the surrounding Cu$^{2+}$ sites. Early theoretical treatment of the problem expected that a nonmagnetic impurity can only induce a local moment when the "mother" phase has a spin gap [10]. Further studies of this problem concluded that in a doped antiferromagnet with no spin gap an induced localized magnetization will exist around the impurity but it does not correspond to a free moment. At low enough temperatures where the staggered correlation length is larger than the localization length associated with the impurity distortion, this localized moment is always aligned with the local staggered order [11]. While the question of whether the magnetic moments exist in a global sense and in particular whether the moments are observable below the superconducting transition temperature is still an open question, new theoretical ideas argue that the moments may be partially screened below the spin gap and fully screened deep in the superconducting state [12] in the cuprates. Recently, there has been some theoretical work done considering the fact of a possible Kondo-screening. In a conventional s-wave superconductor, the Kondo effect is suppressed by the formation of the superconducting gap, as shown by Abrikosov and Gorkov [13]. In d-wave superconductors, however, there are quasiparticles in the node lines which could cause a Kondo screening. Indeed, Fradkin and Cassanello recently calculated the thermodynamic properties of a d-wave superconductors with magnetic impurities [14]. They predicted a screening mechanism analogous to the exchange coupling between magnetic impurities and the electrons in a Fermi liquid that causes the Kondo effect. In the clean limit, the Dirac like quasiparticles and the magnetic impurity form a singlet state. The screening is only effective above a critical exchange coupling , because in contrast to the Kondo effect in metals, the density of states of normal quasiparticles in a d-wave superconductor vanishes at the Fermi surface. In this limit, the magnetic susceptibilty disappears superlinearly, hence the effect can be considered as overscreening.

Observing the effect of a low concentration of Zn doping on the magnetic behavior of the cuprates in the superconducting state is practically impossible due to the strong signal associated with the vortex state. However, applying a magnetic field will split the free Zn-induced moments into a two-level system and a Schottky anomaly should be apparent in the low temperature specific heat. In the present paper we provide a comprehensive study of the effect of Zn impurities on the behavior of low-lying electronic excitations in YBCO. To this end we study the specific heat of pure, 0.15%, 0.31% and 1% Zn-doped single crystals YBCO samples. We extend earlier mea-surements on Zn-doped polycrystalline YBCO samples. Compared to pure YBCO we find that: i) The linear-T term, which has been regularly observed in specific heat measurements of YBCO [15] is greatly increased. This result is only qualitatively consistent with the "dirty d-wave" scenario [16,17] used to explain the penetration depth measurements and the suppression of $T_c$. ii) The $T^3$ term decreases, suggesting either a surprisingly large increase in the Debye temperature or a new electronic contribution with a similar temperature dependence. iii) A low temperature upturn $T^{-2}$ below ∼3.5 K is observed. This upturn is seen in many specific heat measurements [18], though, its origin is still unclear. iv) The most intriguing result of our measurements is that there is no apparent increase in the number of free magnetic moments as indicated by a Schottky anomaly in any of the Zn-doped samples as compared with the pure sample. Normal state measurements of the magnetization of the 1% sample is in agreement with previous measurements indicating a magnetic moment of the size ∼0.36 $\mu_B$/Zn. From these results, we conclude that the magnetic moments that exist in the normal state must be screened at low temperatures in the superconducting state.

Pure and Zn-doped YBa$_2$Cu$_3$O$_{6.95}$ (optimally doped) single crystals were grown by a flux-growth technique. The detailed process and the characterization of the samples is described in [4]. The Zn concentration was determined by comparing $T_c$ of the crystals to powder samples with known Zn concentration and annealed at the same conditions. This method gives an accuracy of 0.01 - 0.02% Zn per Cu. With increasing Zn-doping, the $T_c$ is suppressed by $\Delta T_c$ ∼12K/% Zn. Magnetic measurements showed that in our 1% sample the $T_c$ is suppressed by 12 K. Specific heat data were taken using the relaxation method, which especially allows the measurement of relatively small samples. The addenda heat capacity of the calorimeter was measured independently and subtracted from the sample measurements. A more detailed description is found elsewhere [15,19].

Specific heat data on the Zn-doped YBCO crystal were taken between 2 and 12 K at zero field and between 2 and 8 K at 0.5, 2, 4, 6, and 8 Tesla. The specific heat was measured as a function of temperature at constant magnetic field which was applied above $T_c$ before cooling the sample. The zero-field results are shown in figure 1, plotted as $c(T)/T$ vs. $T^2$ and compared with pure (0% Zn) samples. In this representation, for a normal metal, the slope $\beta$ is the coefficient of the lattice $T^3$ term and the intercept $\gamma$ is the coefficient of the linear electronic heat capacity and proportional to its density of states. Deviations from this $\gamma T + \beta T^3$ behavior are not very noticeable for the nominally pure sample, but with increasing Zn concentration the samples show a low-temperature upturn below ∼3 K. Also, from a visual observation, the curves for the Zn-doped samples have larger $\gamma$, and smaller $\beta$ suggesting higher electronic density of states and higher Debye temperatures with increasing Zn concentration. While the cause of the low-temperature upturn is still unknown, there are some indications that it is related to interaction between paramagnetic impurities and a local field [18,20] or a result of collective behavior of these impurities [18]. This is supported by our finding that this upturn is somewhat influenced by the quality of the sample. Previous reports of early samples of YBCO, showed a much larger low-temperature upturn in the specific heat, thus, a straightforward fitting of the data at higher temperatures (e.g. ≥ 5 K) to avoid the upturn will result in misleading values of both the $\gamma T$ term and the $\beta T^3$ term [20]. Although the "upturn effect" is not too large below 3.5 K, and does not effect our overall data, we still include it in our analysis as discussed below.

The magnetic field dependence is shown in figure 2 for the 1% Zn sample. The lower Zn concentration samples show similar behavior. The most notable result here is the further increase in the extrapolated intercept with the $c(T)/T$ axis, an indication of further increase in the quasi-particles density of states upon the application of magnetic field. Based on our previous results on pure YBCO crystals [15] we first attempt the general fit:

$$c(T, H = 0) = A(0)/T^N + \gamma(0)T + \beta T^3,$$
$$c(T, H \geq 0.5T) = A(H)/T^N + \gamma(H)T + \beta T^3$$
$$+ c_{Sch}(T, H) \qquad (1)$$

where we omitted the $T^2$ term for the Zn doped samples as we no longer expect to have clean lines of nodes, and we attempt to account for the uncertainty in the upturn by allowing the power law used to describe it to vary as $T^{-N}$ where N=1,2. In these equations, $\beta$ is field independent and the $\beta T^3$ term is assumed to result from phonons. $c_{Sch}(T, H)$ is a Schottky anomaly associated with a field-independent molar concentration $\nu_{Sch}$ of free spin-1/2 moments. Finally, we have a free function $\gamma(H)$. For pure samples, the coefficient of this linear-T term is found to obey $\gamma(H) = \gamma(0) + \tilde{\gamma}(H)$ where $\gamma(0)$ decreases as twin boundaries and oxygen vacancies are removed from the crystal (for O$_{6.99}$, $\gamma(0)$ ∼ 1 mJ/mol-K$^2$), and $\tilde{\gamma}(H) \propto H^{1/2}$ with a coefficient which agrees semiquantitatively with the prediction for lines of nodes [21]. For the low temperature upturn, in fitting only the zero-field data, the best fit was obtained with N=2. However, a global fit using data of all fields for a given Zn concentration results in a best fit with N=1. As our results and in particular the conclusions based on the results were not sensitive to the choice of N, we used N=1 in all of the following data analyses. In general we found that the field dependence of the 1/T low temperature upturn is not consistent with increasing spin-1/2 impurities: it does not change dramatically with applied field and does not become a Schottky-like term either. The coefficients that we obtained in this fit are given in Table I.

Starting our analysis with the effect of Zn-doping on $\beta$, we find that $\beta$ = 0.308, 0.284, and 0.274 mJ/mol-K$^4$ for the 0.15%, 0.31% and 1% Zn doped samples respectively, as compared with 0.392± 0.001 mJ/mol-K$^4$ for the nominally pure sample. If this $T^3$ term is only asso-ciated with phonons, it corresponds to an increase in the Debye temperature from 403 K in the nominally pure sample to 454 K in the 1% Zn-doped sample. Such a large increase is very surprising and may indicate that a portion of the $T^3$ term is electronic in nature. The linear term, $\gamma(0)T$, also increases with Zn doping. The zero-field linear-T term increases from 3.0 mJ/mol-K$^2$ in the nominally pure sample to 4.3, 5.8 and 11.7 mJ/mol-K$^2$ in the 0.15%, 0.31% and 1% Zn doped samples respectively. Assuming a normal state linear T coefficient $\gamma_n \sim 20$ mJ/mol-K$^2$, this implies a residual density of states ratios $n_{res} = N(E_F)/N_n$ of 0.066, 0.14, and 0.43 for the three crystals (here $N(E_F)$ is the density of states at the Fermi level and $N_n$ is the density of states in the normal state that results in that $\gamma_n$). From penetration depth measurements of crystals from the same batches [4], an analysis based on a "dirty d-wave" [17] found a scattering parameter $\Gamma/T_c \simeq 0.018-0.030$ for 0.31% Zn doping and $\Gamma/T_c \simeq 0.006-0.009$ for the 0.15% Zn doping. In the limit of strong resonant scattering the dirty d-wave model predicts $\Gamma \propto n_{res}^2$ with a logarithmic correction of order $ln(T_c n_{res}/\Gamma)$. Ref. [17] shows reasonable agreement with that prediction using penetration depth data of 0.15% and 0.31%. To analyze the specific heat we first need to estimate the "Zn-like" impurities in the nominally pure sample. Again, the penetration depth of similar samples give a scattering parameter $\Gamma/T_c$ of about 0.0008 which corresponds to an impurity concentration of 0.013% and a residual density of states ratio of 0.03, which implies a contribution of $\sim$0.6 mJ/mol-K$^2$ to the zero-field linear term. The residual $\sim$2.4 mJ/mol-K$^2$ most probably come from the unfilled chains as was found by measuring the high oxygen content crystals [15]. Using the above data figure 3 shows the change of the zero field linear term from the Zn-free sample ($\Delta\gamma(0)$) as a function of the Zn concentration. For comparison we also show in figure 3 the expected $\Delta\gamma(0)$ expected from "dirty d-wave" theory [17]. While as for the penetration depth data, the low Zn concentration fits the "dirty d-wave" theory reasonably well, stronger deviation appears for the 1% sample indicating a more linear dependence on the Zn concentration. Table I summarizes the results presented above.

For pure samples, the field dependent coefficient of the linear-T term obeys $\gamma(H) = \gamma(0) + \tilde{\gamma}(H)$. While $\tilde{\gamma} \propto H^{1/2}$ for pure samples, Zn doping alters this behavior and the 1% Zn sample showed $\tilde{\gamma} \propto H$ as is shown in figure 4. This suppression of the $H^{1/2}$ term was expected because of increased scattering. In Volovik's work [21], the $H^{1/2}$ magnetic field dependence of the density of states results from the use of the intervortex spacing as a cutoff to the spatial integral over the density of states in the vortex. For the large Zn concentration such as our 1% sample we expect a much shorter cutoff length. In the unitary limit the scattering rate at low frequencies (and temperatures) saturates at a value $\ell \sim v_F n_{res}/2\Gamma$ [17]. Using our results for $\Gamma$ we find that this length is of order 100 Å, thus explaining the linear field dependence.

Finally we turn to the search of the Zn induced magnetic moments. As noted in the introduction the Zn-induced impurities are expected to be observed in finite magnetic field specific heat measurements. For a two-level system this is a Schottky-like anomaly given by:

$$c_{Sch}(T,H) = \nu_{Sch} R \left(\frac{\Delta}{k_B T}\right)^2 \frac{e^{\frac{\Delta}{k_B T}}}{\left(1 + e^{\frac{\Delta}{k_B T}}\right)^2} \qquad (2)$$

where $\Delta$ is the size of the two-level system gap. Clearly a true two level system exists for spin 1/2 impurities only. In the case of Zn impurities we expect an effective moment of order one Bohr-magneton [5], hence, a simplified two-level system with a moment $\mu$ and a gap $\Delta = 2\mu H$ should be a good approximation for the expected efect on the specific heat. Also in equation 2, $\nu_{Sch}$ is the molar concentration of moments and $R$ is the gas constant. Impurity moments are always found in specific heat measurements of high-$T_c$ materials even for "pure" crystals. For optimally doped YBCO a concentration of about 0.05% - 0.1% per copper atom is found [15]. For higher oxygen concentration this number is smaller providing us with evidence that most of these moments come from uncompensated copper atoms in the chains. As mentioned above, it has been argued that Zn impurities induce moments that for the concentration range and oxygen level of our samples is expected to be $\mu = 0.32\mu_B$ to $0.36\mu_B$ [6]. Thus, it is easy to calculate the expected Schottky anomaly associated with the Zn moments; in particular we expect that since the Zn moments originate from the Cu-O planes and the "unaccounted" spin-1/2 moments come from the chains the Zn contribution will simply be an addition to the pure crystal heat capacity. With the above introduction, the most surprising result is that we did not find any change in the Schottky anomaly for any of the Zn doping all the way to 8 Tesla, the maximum field we measured. This result can most dramatically be seen in figure 5 where the specific heat data at 8 Tesla is plotted together with the calculated Schottky anomaly expected for 1% Zn with magnetic moment per Zn of $\mu = 0.32\mu_B$ (the smaller expected contribution). Also plotted in figure 5 is a Schottky anomaly with $\mu = 1.28\mu_B$, displaying the possibility of a "composite" moment that results from local ferromagnetic interaction of moments on four neighboring oxygen atoms [5]. If a significant amount of either $0.32\mu_B$ or $1.28\mu_B$ moments were present, it would be readily apparent in the size of the Schottky anomaly, which is unchanged from undoped samples. Fitting with a Schottky term with $1.28\mu_B$ moments in addition to the $1\mu_B$-spin-1/2 anomaly finds no more than 0.02% additional spins per Cu and is consistent with no $1.28\mu_B$ moments at all. At $0.32\mu_B$ per moment, the presence of 1% spins per Cu would create a Schottky tail which wouldn't roll over at even the highest fields and lowest temperatures measured. However, this upturn would increase proportionally to the square of the applied field, being 16 times larger at 8 Tesla than at 2 Tesla. Assuming that all of the low temperature upturn in the 8 Tesla data is from a small-moment Schottky anomaly, we obtain an upper limit of 0.03% spins per Cu, for $0.32\mu_B$ moment spins. Figure 5 highlights this discrepancy, showing that even with different fitting methods, there are few, if any, of the Zn impurities creating free magnetic moments in the crystal.

The absence of Zn-induced magnetic moments in the superconducting state may be explained in two ways. The first is that the moments exist only in a local sense and thus will be found in local measurements such as Knight shift but will be missed in bulk measurements such as the present specific heat due to the "spread" of the moment throughout the whole sample [11]. To check for this hypothesis we measured the susceptibility of the 1%-Zn doped sample. A weak Curie behavior was found which is of similar magnitude to that measured by Zagoulaev *et al.* [5] implying a similar size of magnetic moment per Zn impurity. The fact that the Zn-moments exist in the normal state may be explained by the fact that we have been measuring a doped Heisenberg antiferromagnet which is close to optimal doping, and for which the spin gap has not collapsed yet.

The second possibility for the "disappearance" of the moments is that they are screened in a Kondo fashion. Such a possibility was recently considered by Nagaosa and Lee [12] and by Cassanello and Fradkin [14]. Relevant to our measurements is the Cassanello and Fradkin work in which they postulated that magnetic impurities in a clean d-wave superconductor will be screened, and show that the screening can be described as a multichannel non-marginal Kondo problem. For simplicity they assume spin-1/2 impurities that are located on the Cu sites in the planes. Using expected values of the exchange coupling, they find a critical temperature $T_K \sim 10$ K, below which the impurities will be overscreened. Their calculation of the low-temperature specific heat yielded the expressions:

$$c(T,H) = 9\zeta(3)\nu_i R N_c \delta^2 \left(\frac{T}{T_K}\right)^2 ln\left(\frac{\Delta_0}{k_B T}\right)$$
$$for\, k_B T > \mu H$$
$$c(T,H) = \left(\frac{\pi^2}{3}\right)\nu_i R N_c \delta^2 \left(\frac{T}{T_K}\right)\left(\frac{\mu_B H}{k_B T_K}\right) ln\left(\frac{\Delta_0}{\mu_B H}\right) \quad (3)$$
$$for\, k_B T < \mu H$$

Here $\nu_i$ is the impurity concentration per mole, R is the universal gas constant, $\zeta(3)$ is the Rieman zeta function of argument 3, $N_c$ is the rank of the symmetry group of the impurity spin, taken to be 2 in the relevant case (see [14] for details), $\delta$ is a dimensionless parameter related to the strength of the screening ( the singlet amplitude), $\Delta_0$ is the maximum gap of the $d_{x^2-y^2}$ superconductor, and $T_K$ is the Kondo temperature associated with the impurities. Before presenting our fit to the Cassanello and Fradkin's calculation we note that the calculation are based on a very simple model that assumes that the quasiparticles that screen the moments come from the nodes only. However, Zn impurities cause the depression of the order parameter close to the impurity site due to potential scattering (i.e. pair breaking). This effect is the main cause of the depression of $T_c$, and was not included in the calculations. However, all the effects related to Zn scattering are local in nature and thus should reduce to a generalized Andreev boundary condition (generalized because of the different symmetries that may be involved, including inter-node scattering and so on). It is very unlikely that these effects will help in the Kondo screening since these potentials are invariant under spin rotations [22]. With the above discussion in mind we fitted our Zn-doped specific heat data with the screening model, i.e. equations 3. Since the model provides us with the limiting cases only for temperature and fields we first fit the high field, assuming that all the field dependence comes from the screening. We then introduce an interpolation function that is exponential in nature which connects the low and high magnetic fields. We show the results of the fit in figure 6 while the fitting parameters are shown in TABLE II. Again, compared with Zn-free sample the Schottky anomaly does not change, the $T^3$ contribution is smaller and a small low temperature upturn is observed. For the 1%Zn we find $(\delta/T_K) = 0.0082$, a value within the range discussed in [14]. For that sample we also find that the quality of the fit degrades at high fields, possibly due to the model becoming invalid with the relatively large DOS at the nodes and its simultaneous smearing by the Zn doping and the magnetic field. For the 0.31%Zn the value $(\delta/T_K) = 0.003$ seems somewhat low as we expected to first approximation that the two values will be the same.

In conclusion, we presented in this paper a comprehensive study of the low temperature heat capacity of Zn-doped YBCO single crystals. Several trends have been found as a function of Zn doping. The $T^3$ term decreases with increasing Zn while the liner term increasing, indicating an increasing DOS at the nodes. This residual density of states is only qualitatively consistent with a d-wave resonant scattering theory. The 1%Zn clearly deviates from the theoretical expectation, showing excess DOS. A low temperature upturn is consistently observed for the Zn-doped samples, similar to early measurements on more disordered YBCO samples. The cause and exact form of this upturn is still not clear. The most surprising result of our study tough is that no Schottky anomaly is observed for the presumed magnetic moments created by Zn doping. Based on a recent theory by Cassanello and Fradkin we fitted the data with Kondo screened moments. The relatively good quality of the fit and the reasonable parameters extracted from it give us confidence that indeed Zn-moments in YBCO crystals are screened by quasiparticles in the nodes of the $d_{x^2-y^2}$ superconductor.

*Acknowledgments:* Work at Stanford University was supported by Air Force Office of scientific Research. S.G.D. was supported by a Feodor-Lynen-Fellowship of the Alexander von Humboldt-Stiftung. The authors wish to thank Kathryn Moler for her valuable contributions at the initial stages of this work. we also wish to thank Doug Scalapino and Eduardo Fradkin for many useful discussions.

FIG. 1. The effect of Zn doping on the zero-field specific heat of $YBa_2Cu_3O_{6.95}$.

FIG. 2. The temperature dependence of the specific heat (plotted as c(T)/T of 1% Zn doped YBCO, cooled in applied fields from 0 to 8 Tesla.

FIG. 3. The zero-field $\gamma$ term as a function of Zn doping. Line that connects the points is guide to the eye, emphasizing the sub-linear trend of the data.

FIG. 4. The field dependence of the coefficient of the linear-T term, $\gamma(H)$, as determined from the global fits described in the text. Note, a $T^2$ term was allowed for the nominally pure sample. Connecting lines are only guide to the eye.

FIG. 5. The Schottky effect observed in the 1%Zn-doped sample together with an expected Schottky effect associated with 1% moments of size: 0.32 $\mu_B$ (dashed line) or 1.28 $\mu_B$ (dash-dotted line), see text. solid line through the 8 T data is the fit as presented in figure 6. Solid circles are the zero-field data.

FIG. 6. Fit of the screening approximation [14] to the 1% Zn doped sample. The fit assumes that all the field dependence in that limit comes from the screening.

TABLE I. Parameters for fit with equation 1 (no screening term).

| Zn % | $T_c$ [K] | $\beta$ [mJ/mol-K$^4$] | $\nu_{Sch}$R [mJ/mol-K] | A(0) [mJ/mol] | $\gamma(0)$ [mJ/mol-K$^2$] |
|---|---|---|---|---|---|
| 0.0 | 93 | 0.392±0.001 | 24±1 | no term | 3.0±0.1 |
| 0.15 | 90 | 0.308±0.003 | 20±2 | 7±1 | 4.3±0.1 |
| 0.31 | 88 | 0.284±0.003 | 25±2 | 16±2 | 5.8±0.2 |
| 1.0 | 81 | 0.274±0.003 | 24±2 | 3±1 | 11.7±0.2 |

TABLE II. Parameters for fit with equation 3 (including screening term).

| Zn % | $\beta$ [mJ/mol-K$^4$] | $\nu_{Sch}$R [mJ/mol-K] | $\gamma(0)$ [mJ/mol-K$^2$] | $\delta/T_K$ [K$^{-1}$] |
|---|---|---|---|---|
| 0.31 | 0.284±0.003 | 25±2 | 5.8±0.2 | 0.003±0.002 |
| 1.0 | 0.274±0.003 | 24±2 | 11.7±0.1 | 0.0082±0.0004 |

Figure 1

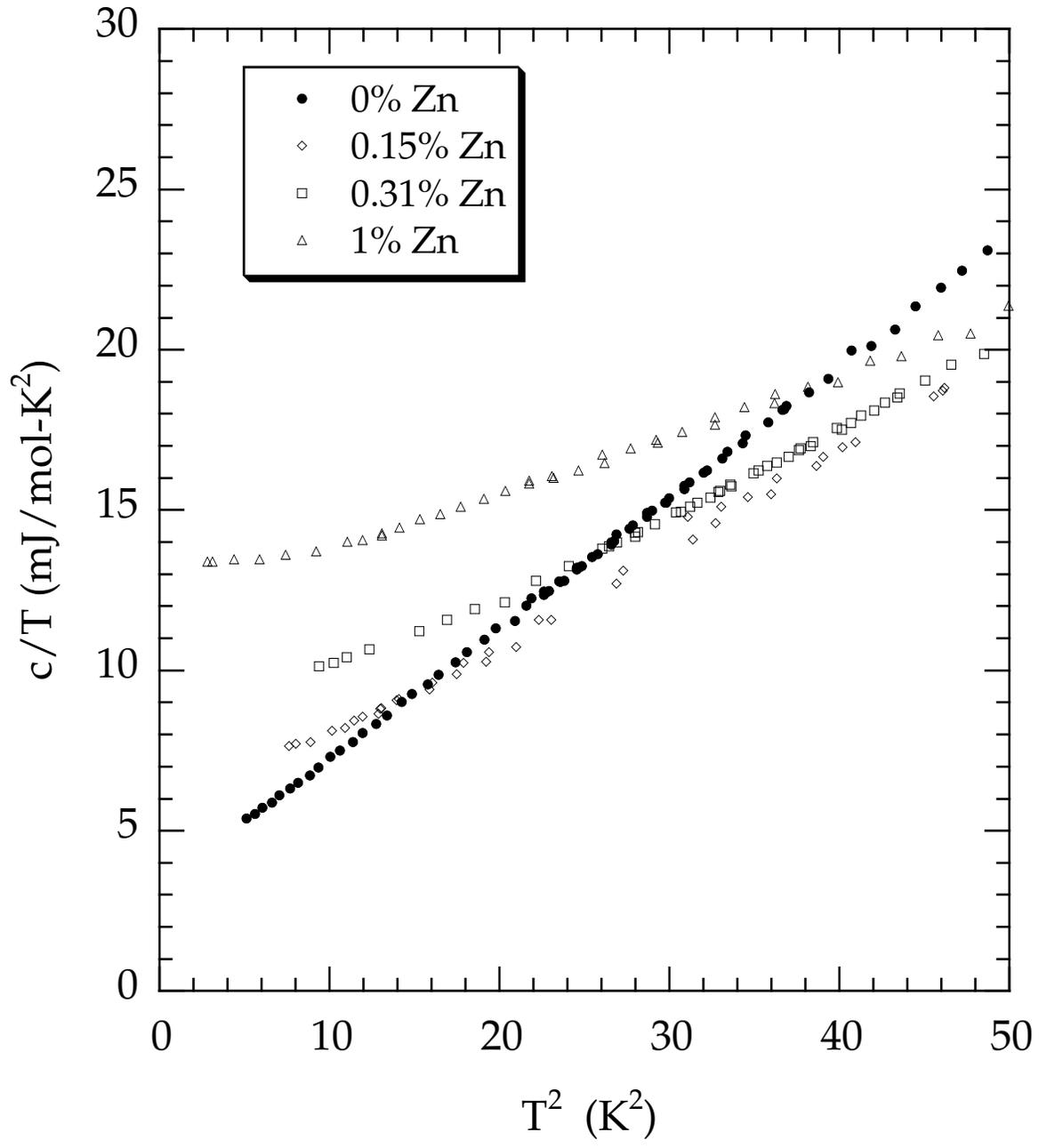

Figure 2

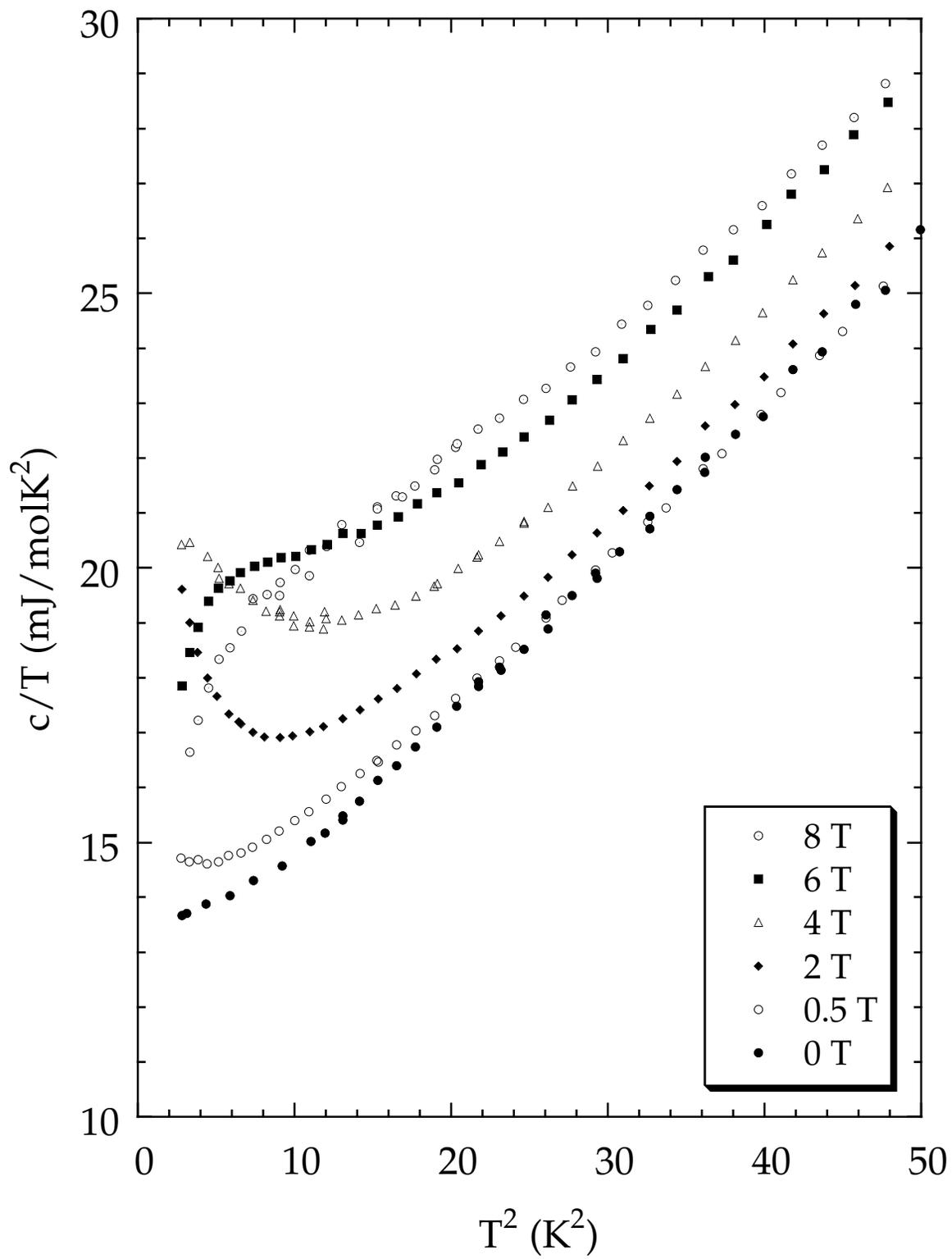

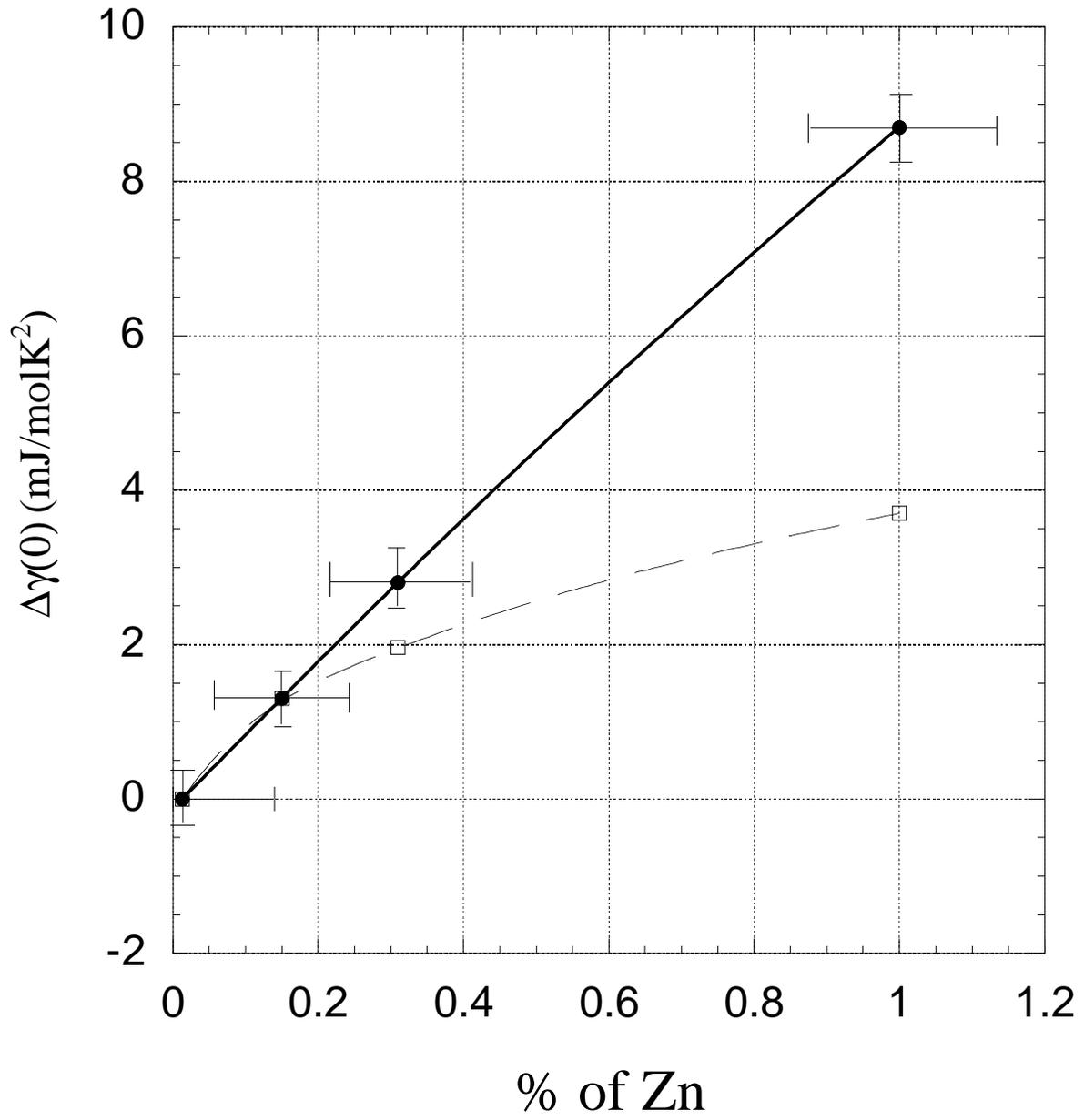

Figure 3

Figure 4

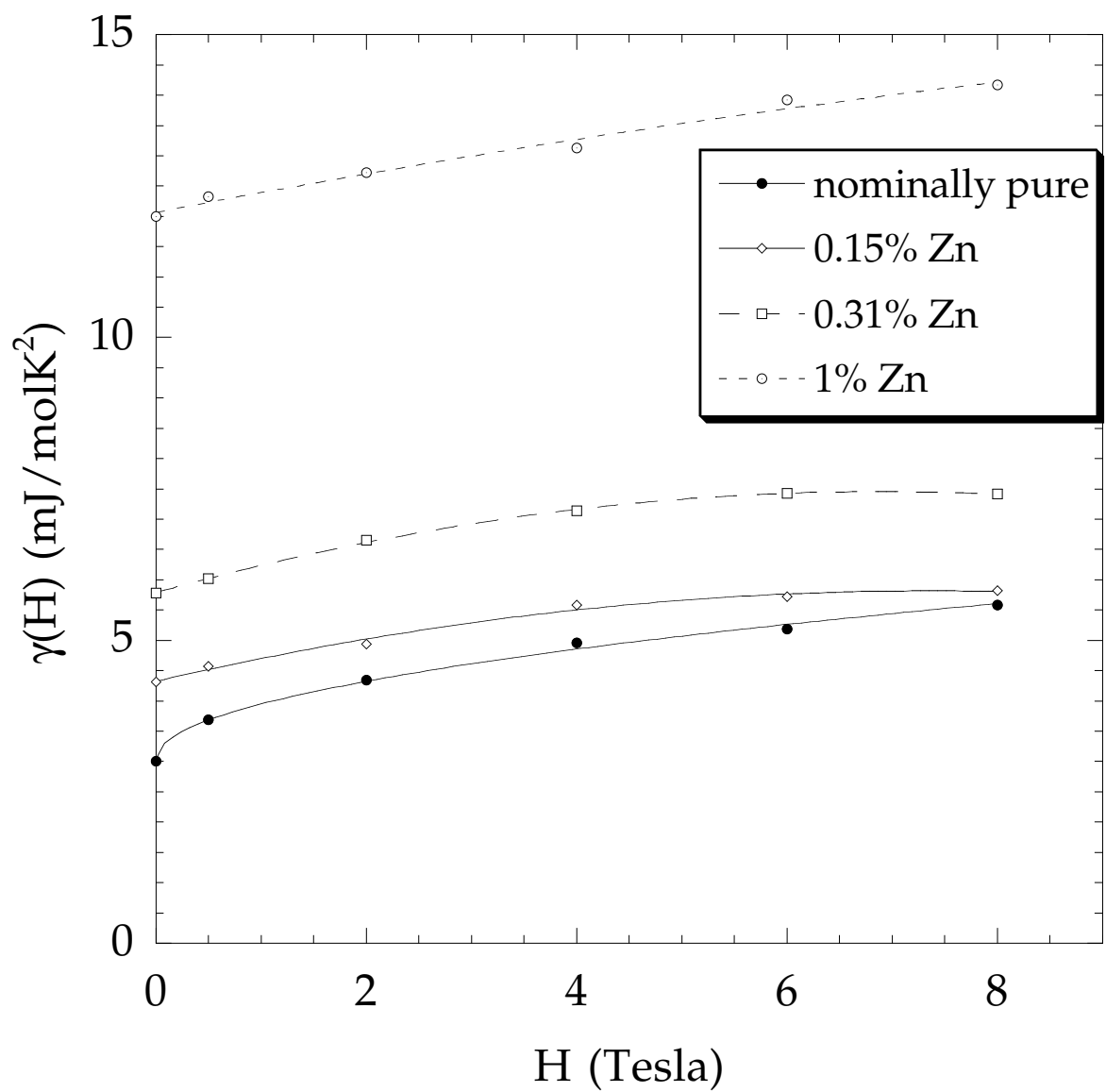

Figure 5

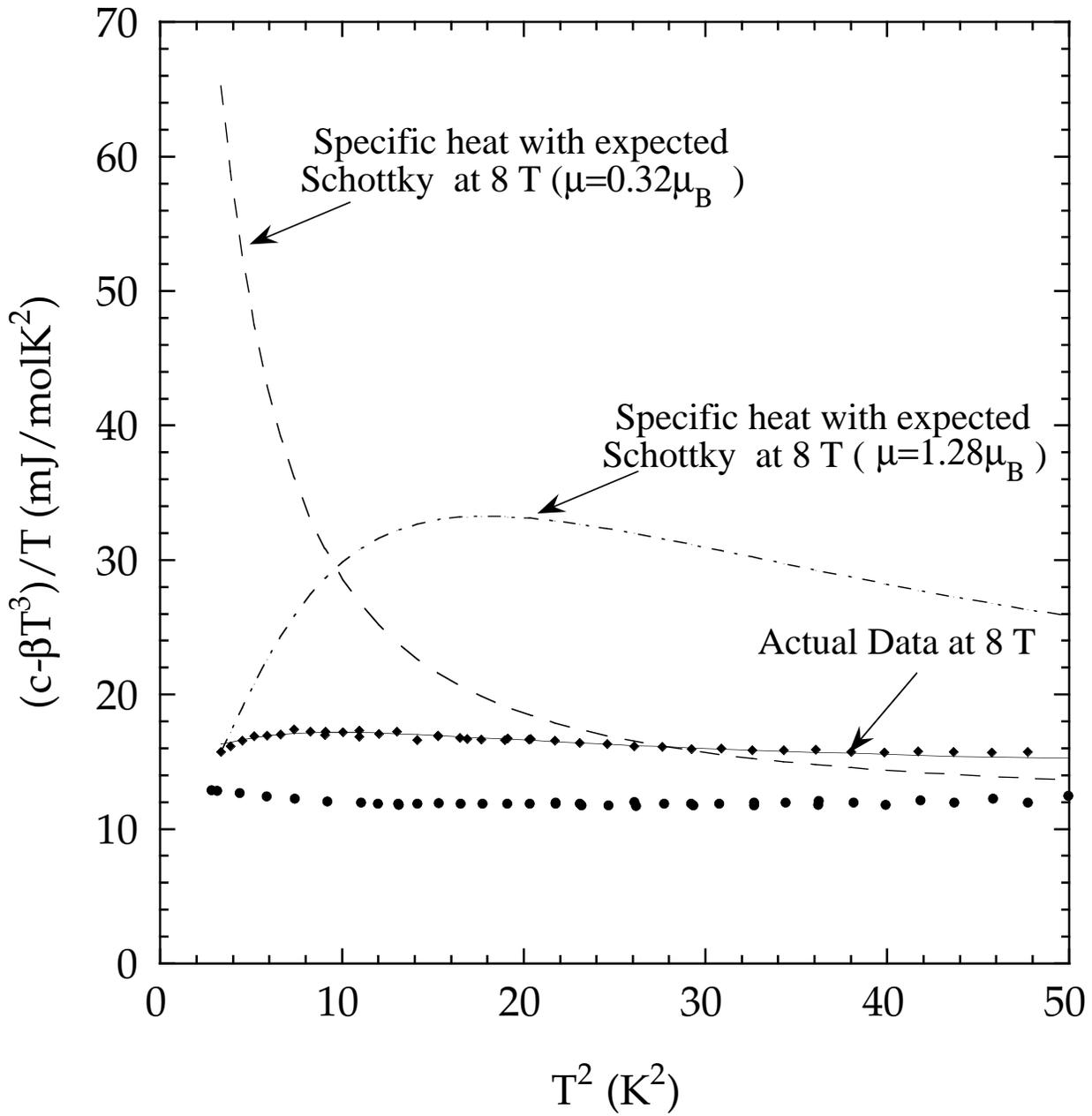

Figure 6

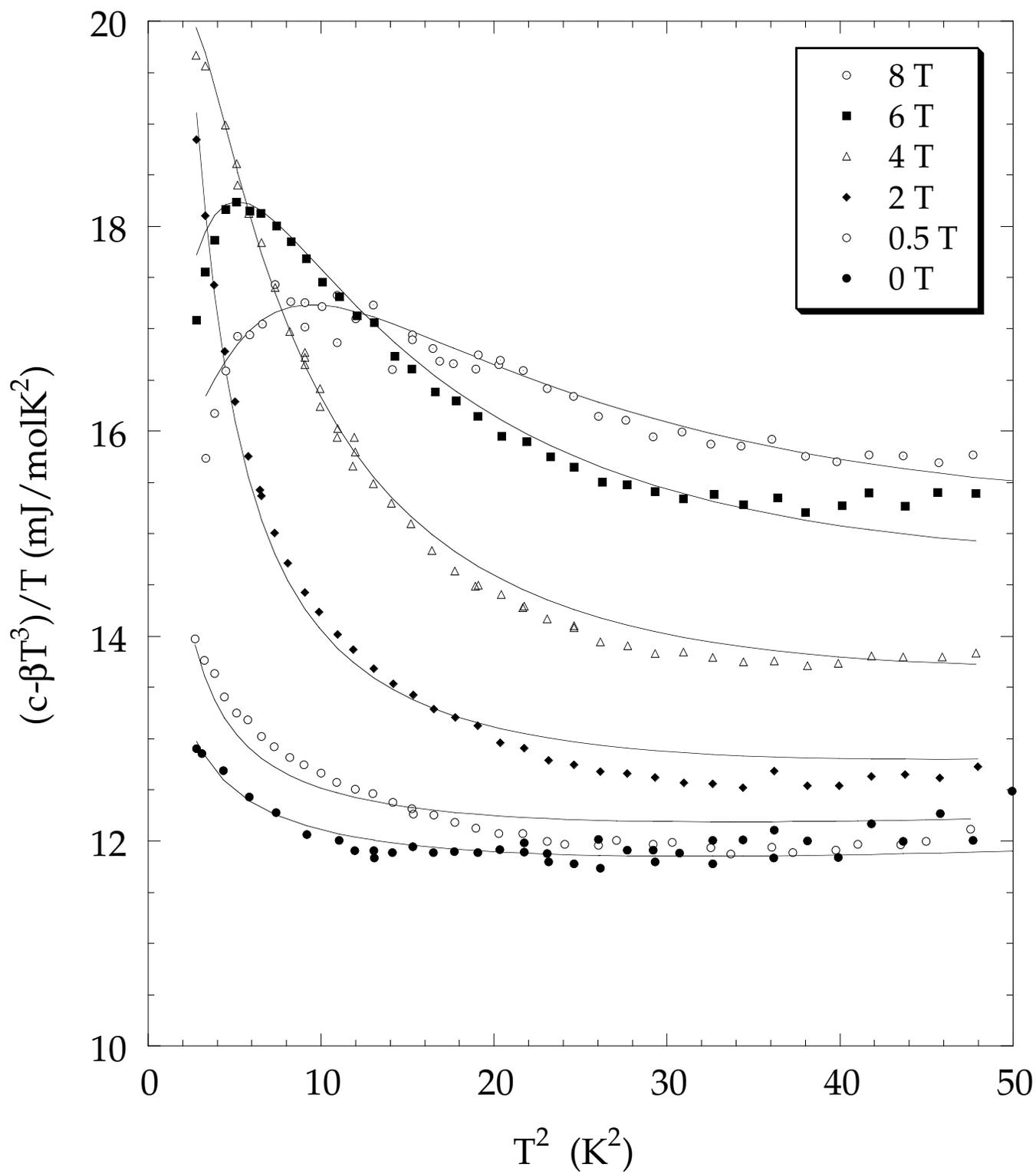